\documentclass[]{pasj00}
\begin{document}
\SetRunningHead{F. Oktariani and A.\,T. Okazaki}
{Global Disk Oscillations in Binary Be Stars}
\Received{2008//}
\Accepted{2008/??/??}

\title{Global Disk Oscillations in Binary Be Stars}

\author{Finny \textsc{oktariani}} 
\affil{ Department of Cosmoscience, Graduate School of Science, Hokkaido University, Kita-ku, Sapporo 060-0810, Japan}
\email{finny@astro1.sci.hokudai.ac.jp}
\and
\author{Atsuo T. \textsc{okazaki}}
\affil{Faculty of Engineering, Hokkai-Gakuen University, Toyohira-ku, Sapporo 062-8605, Japan }

\maketitle

\begin{abstract}
We study the effects of the tidal interaction with the companion, via
orbital separation and binary mass ratio, on the global one-armed oscillation
modes in disks around binary Be stars. Our model 
takes into account the three-dimensional effect that contributes to the mode
confinement, which was recently found by \citet{Ogi08}. We
find that the one-armed oscillations are well confined in systems with disks larger
than a few tens of stellar radii. In such systems, the oscillation period depends little on the binary parameters. On the other hand, in systems with smaller
disks, where the mode confinement is incomplete, the oscillation
period increases with increasing orbital separation and/or 
decreasing binary mass ratio. The eigenmode is insensitive to
the spectral type of the central star.
Our results suggest that the dependence of V/R
oscillation period on the orbital separation and binary mass
ratio should be observed only in short period binary systems, 
and that, for systems with a similar orbital
period, those with higher mass ratios will show shorter V/R variations.
\end{abstract}

\section{Introduction}
\label{sec:intro}
Be stars are non-supergiant B-type
stars whose spectra show, or have at some time shown, one or more
Balmer lines in emission (\cite{Col87}). A Be star is a rapidly rotating, B-type star as the central star with a 
two-component circumstellar envelope, a polar wind and an
equatorial disk. The polar wind is a low-density, fast outflow emitting UV radiation. The wind
structure is well explained by the line-driven wind model
(\cite{CAK75}; \cite{FO86}). On the other hand, the equatorial disk
is a geometrically thin, high-density plasma in nearly
Keplerian rotation (e.g., \cite{Por03}),
from which the optical emission lines and
the IR excess arise. Although there are still
several competing scenarios for the Be disk formation, the most
promising one is the scenario where the disk is formed by viscous
decretion of gas ejected from the central star (\cite{Lee91}; see also
\cite{Por03}, and references therein).

Many Be stars show long-term variations in their spectra over
years to decades. One of these phenomena is called long-term V/R
variations, which are variations of the ratio of relative
intensity of violet (V) and red (R) peaks of a double-peaked
emission line profile. The period of the long-term V/R variations is
typically in the range 5-10\,yr for isolated Be stars. 
It is widely accepted that the long-term V/R variations are attributed
to global one-armed (i.e., the azimuthal wave number $m=1$) oscillations
in the Be disk (e.g., \cite{Por03}). It is based on the fact that in a nearly Keplerian disk a one-armed mode shows up as a very slowly revolving perturbation pattern
 (e.g., \cite{Kat83}). The early versions of the one-armed
oscillation model (\authorcite{Oka91} \yearcite{Oka91}, \yearcite{Oka97}; 
\cite{Pap92, Sav93})
qualitatively well explained the observed characteristics of the V/R variations.
However, owing to the lack of the mechanism for confining the modes 
to the inner part of the disk and the high sensitivity of the oscillation period
on various parameters, the model had little predictive power \citep{FH06}.

Recently, the model was advanced greatly. \citet{Ogi08} has solved 
the mode confinement problem by investigating a
three-dimensional effect on the mode characteristics. While 
previous versions of the model had assumed motions to be 
independent of height
since the disk is geometrically thin, he argued that the
variation of the vertical gravitational acceleration around an
elliptical orbit excites an oscillatory vertical motion in an
eccentric disk that should not be neglected. \citet{Ogi08} showed that
the three-dimensional effect alone allows confined prograde
modes. 

Hitherto, theoretical studies of one-armed oscillation modes in Be
disks are mostly limited to isolated Be stars. However, according to
the current census, the fraction of Be stars found in binary systems
is $\sim 1/3$ (\cite{Por03}). The presence of the companion is most
likely to change some features of the global one-armed oscillations in
Be disks. Actually, in Be/X-ray binaries, which consist of a
neutron star and an early-type Be star, the time-scale of V/R
variations is frequently about 1\,yr, which is much shorter than the
typical V/R time-scale for isolated Be stars (e.g., \cite{Rei05}).

Moreover, there are observationally two types of long-term V/R
variations in binary Be stars (\cite{Ste07}). One is quasi-periodic
V/R variations, where V/R peak ratio varies quasi periodically, which
shows no correlation with their orbital phase. This type
of V/R variations are similar to those usually observed for V/R
variations in isolated Be stars. The other type is periodic V/R
variations locked to the orbital period. This characteristic is found
only in binary Be star systems. Unfortunately, no theoretical 
explanation is available yet for the phase-locked V/R variations. 

In this paper, we investigate the effects of the companion on the
characteristics of one-armed oscillation modes in binary Be stars,
taking into account the tidal effects in addition to the
three-dimensional effect. For simplicity, we assume the binary orbit to be circular.
We find that the mode is well confined in disks larger than
a few tens of stellar radii, which is consistent with 
\citet{Ogi08}. In smaller disks, however, the mode confinement is
incomplete and the oscillation period depends on the binary parameters significantly.

The structure of this paper is as follows. In section 2, we
describe the unperturbed disk and derive the basic
equations for linear, one-armed perturbations, following
the formulation by \citet{Ogi08}. The boundary
conditions to be adopted are also discussed. In section 3, we present our numerical
results, and in section 4, we compare our results with the
observational features of V/R variations in binary Be
stars. The final section is devoted to conclusions.

\section{Basic equations for $m=1$ density waves}
\subsection{Basic equations}
As an unperturbed state, we take a geometrically thin, axisymmetric disk that locally rotates at a nearly Keplerian speed 
and is in hydrostatic equilibrium in the vertical direction. 
Because of the strong photoionization heating from the stellar radiation, 
this Be disk is taken to be isothermal at $0.6 T_\mathrm{eff}$ 
\citep{Car06}, 
where $T_\mathrm{eff}$ is the effective temperature of the central star. 
For simplicity, we assume the binary orbit to be circular and
neglect the advective motion and viscous effects in the Be disk. 

We use the cylindrical coordinate system $(r, \phi, z)$. The equation of continuity and the equation of motion are respectively given by
\begin{equation}
\frac{\partial\rho}{\partial t}+ \boldsymbol{\nabla\cdot}\left(\rho \boldsymbol{v} \right)= 0
\end{equation}
and
\begin{equation}
\left[\frac{\partial}{\partial t}+ \left(\boldsymbol{v \cdot \nabla}\right)\right]\boldsymbol{v} = -{\bf\nabla}\Psi - {\bf\nabla} h.
\end{equation}
Here $\boldsymbol{v}$ is the velocity vector, $\Psi$ is the gravitational potential,
and $h = c_\mathrm{s}^2 \ln \rho$ is the enthalpy of an isothermal gas, where $c_\mathrm{s}$
is the isothermal sound speed and $\rho$ is the density.

As far as eigenmode oscillations are concerned, we may consider only the azimuthally averaged 
tidal potential \citep{Hir93}.  We also take into account the rotational deformation 
of the rapidly rotating Be star, including only monopole and quadrupole term \citep{Pap92}. 
The potential in the disk midplane, $\Psi_\mathrm{m}$, is then reduced to
\begin{eqnarray}
\Psi_\mathrm{m} & \simeq & -\frac{GM_{1}}{r} \left[ 1 + k_{2}\left(\frac{\Omega_{1}}{\Omega_\mathrm{c}}
 \right)^{2} \left(\frac{r}{R_{1}}\right)^{-2} \right] \nonumber\\ 
 && - \frac{GM_{2}}{D}\left[ 1 + \frac{1}{4}{\left(\frac{r}{D}\right)}^2\right],
\label{eq:potential}
\end{eqnarray}
where $M_{1}$, $R_{1}$ and $\Omega_{1}$ are the mass, radius and angular rotation speed of the Be star, $k_{2}$ and $\Omega_\mathrm{c}=\sqrt{GM_{1}/R_{1}^{3}}$ are its apsidal motion constant and critical angular rotation speed, $M_2$ is the mass of the companion, and $D$ is the binary separation.
In equation~(\ref{eq:potential}), the first term is the gravitational potential of the Be star, in which the quadrupole contribution due to the rotational distortion of the star is taken into account, and the second term is the azimuthally averaged tidal potential.

In circular binaries with small mass ratios ($M_{2}/M_{1} \lesssim 0.3$),
a viscous disk is truncated at the tidal radius given by
\begin{equation}
R_\mathrm{tides} \sim 0.9 R_\mathrm{L}
\label{eq:rtides}
\end{equation}
\citep{WK91}. Here $R_\mathrm{L}$ is the Roche lobe radius of the Be star 
given approximately by
\begin{equation}
R_\mathrm{L} = D \frac{0.49q^{-2/3}}{0.69q^{-2/3} + \ln (1+q^{-1/3})}
\label{eq:roche}
\end{equation}
\citep{Egg83}, where $q = M_{2}/M_{1}$ is the binary mass ratio.
In this paper, we thus assume that the Be disk is truncated at $r=R_\mathrm{tides}$.

The unperturbed equilibrium state of the isothermal disk is given by 
\begin{equation}
\rho = \rho_{m}(r) \exp \left( - \frac{z^{2}}{2H^{2}}\right),
\end{equation}
\begin{equation}
\boldsymbol{v} = [0, r\Omega(r),0],
\end{equation} 
\begin{equation}
p = c_\mathrm{s}^2 \rho,
\label{eq:perturb-iso}
\end{equation}
and 
\begin{equation}
r\Omega^{2}=  \frac{d\Psi_\mathrm{m}}{dr} + c_{s}^{2}\frac{d\ln\rho_\mathrm{m}}{dr},
\label{eq:centrifugal}
\end{equation}
where $\Omega(r)$ is the angular frequency of disk rotation, 
$\rho_{m} (r)$ is the midplane density, and $H(r) = c_\mathrm{s}/\Omega_\mathrm{K}$ 
with $\Omega_\mathrm{K}=\sqrt{GM_{1}/r^{3}}$ is the scale-height of the disk.

From equations (\ref{eq:potential}) and (\ref{eq:centrifugal}), 
we have the explicit form of $\Omega$ as follows{.}
\begin{eqnarray}
\Omega  &=&   \left( \frac{GM_{1}}{r^{3}} \right)^{1/2} 
                 \left\{1 - \frac{q}{2} \left( \frac{r}{D} \right)^{3} 
                  + k_{2} f^{2} \left( \frac{R_{1}}{r} \right)^{2} \right. \nonumber\\
                  & &  \left. + \left( \frac{d \ln \rho_{m}}{ d \ln r} \right) \left( \frac{H}{r} \right)^{2} \right\}^{1/2},
\label{eq:Omegas}
\end{eqnarray}
where $f = \Omega_{1} / \Omega_{c}$.  
Then, the associated local epicyclic frequency $\kappa(r)$ is
explicitly written as
\begin{eqnarray}
\kappa & = & \left[ 2 \Omega \left( 2\Omega + r \frac{d \Omega}{dr} \right) \right]^{1/2}
\nonumber \\
      & = & \left( \frac{GM_{1}}{r^{3}} \right)^{1/2}
           \left\{ 1 - 2q \left(\frac{r}{D} \right)^{3} 
            - k_{2} f^{2} \left( \frac{R_{1}}{r} \right)^{2} \right. \nonumber\\
      && \left. + \left[ 2 \left( \frac{d \ln \rho_{m}}{ d \ln r} \right) 
         + \frac{d^{2} \ln \rho_{m}}{d \ln r^{2}} \right] \left( \frac{H}{r} \right)^{2} \right\}^{1/2}.
\label{eq:kappas}	
\end{eqnarray}

On the above unperturbed {state,} we superpose a linear{, $m=1$} isothermal perturbation {in the form of normal mode
of frequency $\omega$}, which var{ies} as 
$\exp[i(\phi-\omega t )]$. 
The {linearized} perturbed equations {are then obtained} as follows{.}
\begin{equation}
i(\Omega - \omega) v_{r}^{\prime} - 2\Omega v_{\phi}^{\prime} = - \frac{\partial h^{\prime}}{\partial r},
\end{equation}
\begin{equation}
i(\Omega - \omega) v_{\phi}^{\prime} + \frac{\kappa^{2}}{2\Omega} v_{r}^{\prime} = - \frac{ih^{\prime}}{r},
\end{equation}
\begin{equation}
i(\Omega - \omega) v_{z}^{\prime} = - \frac{\partial h^{\prime}}{\partial z},
\end{equation}
\begin{eqnarray}
i(\Omega - \omega) h^{\prime} & +& v_{r}^{\prime} \frac{\partial h}{\partial r} + v_{z}^{\prime} \frac{\partial h}{\partial z}  \nonumber \\
                                         &  & = - c_{s}^{2}  \left[ \frac{1}{r} \frac{\partial (r v_{r}^{\prime})}{\partial r} + \frac{iv_{\phi}^{\prime}}{r} + \frac{\partial v_{z}^{\prime}}{\partial z} \right],
\end{eqnarray}
where ($v_{r}^{\prime}, v_{\phi}^{\prime}, v_{z}^{\prime}$) and $h^{\prime}$ are the perturbed velocity 
and enthalpy, respectively.

{In order to take the three-dimensional effect into account,}
we expand perturbed quantities in the $z$-direction in terms of Hermite polynomials {as}
\begin{equation}
v_{r}^{\prime} (r,z) = \sum_{n} u_{n}(r)H_{n}(\zeta),
\end{equation}
\begin{equation}
v_{\phi }^{\prime} (r,z) = \sum_{n} v_{n}(r)H_{n}(\zeta),
\end{equation}
\begin{equation}
v_{z}^{\prime} (r,z) = \sum_{n} w_{n}(r)H_{n-1}(\zeta),
\end{equation}
\begin{equation}
h^{\prime} (r,z) = \sum_{n} h_{n}(r)H_{n}(\zeta)
\end{equation}
{(\cite{Ogi08}; see also \cite{Oka87}),
where $\mathit{H}_n(\zeta)$} is the Hermite polynomial defined by 
\begin{equation}
H_{n}(\zeta) = \exp \left(\frac{\zeta^{2}}{2} \right) 
   \left(-\frac{d}{d \zeta} \right)^{n} \exp \left(-\frac{\zeta^{2}}{2} \right) 
\end{equation}  
{with} $\zeta = z/H$ {being} a dimensionless vertical coordinate and $n = 0,1,2,\ldots$.

The system of resulting equations is not closed because the equation for $u_{n}$ refers to the higher order term $h_{n+2}$,
which depends on $u_{n+2}$, the equation for which in turn refers to $h_{n+4}$, and so forth. Following \citet{Ogi08}, we close the system of resulting equations by 
assuming $u_{n}\equiv 0$ for $n\geq 2$. Neglecting $u_{2}$ compared to $u_{0}$ is equaivalent to assuming that the eccentricity of perturbed orbit of each gas particle is independent of z. 
If $h_{2}$ is also neglected, then we will obtain the two-dimensional equations. 
{T}he basic equations for linear $m=1$ perturbations in inviscid disks are {then}
given as follows{.}
\begin{equation}
2i\left( \omega_\mathrm{pr} - \omega \right) u_{0} = -\frac{d h_{0}}{dr} - \frac{2h_{0}}{r} + \frac{3h_{2}}{r},
\label{eq:A6}
\end{equation}
\begin{equation}
i\Omega \frac{h_{0}}{c_\mathrm{s}^{2}} - \frac{u_{0}}{2r} + \frac{1}{r \Sigma} \frac{d \left(r \Sigma  u_{0} \right)}{dr} = 0,
\label{eq:A7}
\end{equation}
\begin{equation}
-i\Omega \frac{h_{2}}{c_\mathrm{s}^{2}} + \frac{3 u_{0}}{2r} = 0,
\label{eq:A8}
\end{equation}
where $\Sigma = (2\pi)^{1/2} \rho_{m} H$ is the surface density 
{and $\omega_\mathrm{pr}$ is the local apsidal precession frequency
given by}
\begin{eqnarray}
   \omega_\mathrm{pr} &=& \Omega - \kappa \nonumber \\
   &\simeq & \left( \frac{GM_{1}}{r^{3}} \right)^{1/2}  \left\{ \frac{3}{4}q \left(\frac{r}{D} \right)^{3} + k_{2} f^{2} \left( \frac{R_{1}}{r} \right)^{2}\right. \nonumber\\
   && \left. - \frac{1}{2} \left[  \left( \frac{d \ln \rho_{m}}{ d \ln r} \right) 
         +\frac{d^{2} \ln \rho_{m}}{d \ln r^{2}} \right] \left( \frac{H}{r} \right)^{2} \right\}.
\label{eq:omega-pr}
\end{eqnarray}
{In deriving equations (\ref{eq:A6})-(\ref{eq:A8}), 
we have used approximations $|\omega| \ll \Omega$,
$|\Omega-\kappa| \ll \Omega$, and $(c_{\rm s}/r\Omega)^{2} \ll 1$.
Note that these equations are essentially the same as 
equations (A6)-(A8) of \citet{Ogi08}, except that
our equations implicitly depend on the tidal effect through $\Omega$ and $\kappa$.}

{Eliminating $h_{2}$} from equations (\ref{eq:A6}) and (\ref{eq:A8}), we have 
\begin{equation}
\frac{d h_{0}}{dr} = - \frac{2}{r} h_{0} + \left[ 2 \left(\omega - \omega_\mathrm{pr} \right) - \frac{9c_{s}^{2}}{2 r^{2} \Omega} \right] i u_{0},
\label{eq:A9}
\end{equation}
and equation (\ref{eq:A7}) {is written as} 
\begin{equation}
i \frac{d u_{0}}{dr}= \frac{\Omega}{c_\mathrm{s}^{2}} h_{0} - \left( \frac{1}{2} + \frac{d \ln \Sigma}{d \ln r} \right) \frac{i u_{0}}{r}.
\label{eq:A10}
\end{equation}

We define $Y_{1}$ and $Y_{2}$  as $Y_{1}= u_{0}$ and $Y_{2}= -ih_{0}$. Then the basic equations to be solved are {given by}
\begin{equation}
\frac{dY_{1}}{dr} = \frac{\beta}{r} Y_{1} + \frac{\Omega}{c_\mathrm{s}^{2}} Y_{2}
\label{eq:basic-1}
\end{equation}
and
\begin{equation}
\frac{dY_{2}}{dr} = 2\left[\omega-\left(\omega_\mathrm{pr} + \frac{9c_\mathrm{s}^{2}}{4r^{2}\Omega}\right)  \right] Y_{1}-\frac{2}{r} Y_{2},
\label{eq:basic-2}
\end{equation}
where $\beta=-1/2- \left(d\ln \Sigma/d\ln r \right)$. 
{Note that as shown by \citet{Ogi08}, the term $9c_\mathrm{s}^{2}/4r^{2}\Omega$ 
results from the three-dimensional effect and provides an important contribution to the confinement of the $m=1$ modes.}

\subsection{Boundary conditions}
\label{sec:bc}
We now {consider} the boundary conditions for equations (\ref{eq:basic-1}) and (\ref{eq:basic-2}). 
{In Be stars with growing or persistent equatorial disks, material is likely} 
{in}jected continuously from the star to the disk, so there is no gap between the star and the disk. Since the pressure scale-height of the Be star near the interface is much smaller than that of the disk, the oscillations cannot penetrate the stellar surface. Hence, {throughout this paper,} 
we take the rigid wall boundary condition at the disk inner radius, 
i.e., $v_r^{\prime} = 0$ at $r=R_{1}${, except in section \ref{sec:freebc} 
where we study the effect of the free inner boundary condition}. 
With $Y_{1}=u_{0}$, this condition is written as
\begin{equation}
Y_{1} = 0 \mbox{\hspace{1em} at \hspace{1em}} r=R_{1}.
\label{eq:bc-inr}
\end{equation}

In our model the Be disk is truncated at the tidal radius. Hence, we take the free boundary condition at the outer boundary. This condition is written as
\begin {equation}
 \Delta p = 0,
\label{eq:deltap-0}
\end{equation}
{where $\Delta p$ is the Lagrangian perturbation of pressure.}
{Given that the Eulerian perturbation of pressure, $p^{\prime}$, varies as
$\exp[i(\phi - \omega t )] $, equation~(\ref{eq:deltap-0}) can be written as} 
\begin{equation}
i(\Omega - \omega ) p^{\prime} + v_{r}^{\prime} \frac{\partial p}{\partial r} + v_{z}^{\prime} \frac{\partial p}{\partial z}  = 0.
\end{equation}
Using the same procedure as {in the previous subsection}, 
we have 
 
\begin{equation}
i(\Omega - \omega ) h_{0} + u_{0} \frac{c_\mathrm{s}^{2}}{\rho} \frac{\partial \rho}{\partial r} + u_{0} c_\mathrm{s}^{2} \frac{\partial \ln H}{\partial r}  = 0,
\end{equation}
where {we have used} {\bf $h^{\prime}  = p / \rho $}.
Finally, {with} $Y_{1}=u_{0}$ and $Y_{2}=-ih_{0}$, we have {the outer boundary condition as}
\begin{equation}
\left(\Omega - \omega\right)Y_{2} -\frac{d\ln\Sigma}{d\ln r} \frac{c_\mathrm{s}^{2}}{r} Y_{1}=0 \mbox{\hspace{1em} at \hspace{1em}} r = R_\mathrm{tides}.
\label{eq:bc-out}
\end{equation}

\section{Numerical results}

We solve basic equations (\ref{eq:basic-1}) and (\ref{eq:basic-2}) 
for a range of parameters, including binary separation \emph{D} ($10R_1\leq D\leq 50R_1$) 
and binary mass ratio \emph{q} ($q=0.1$ and $0.3$){,} with the boundary conditions described {above}. In the following calculations, we take a B0V star with $M_{1}=17.5M_{\odot}$, $R_{1}=7.4R_{\odot}$, and $T_{\rm eff}=30000 {\rm K}$  \citep{Cox00} as the Be star model{, except in section~\ref{sec:b5v} where we consider 
a B5V star to study the spectral dependence of the mode characteristics.}
{As for the quadrupole parameter, we take
$k_{2}f^{2} = 5 \times 10^{-3}$ as a representative value,
although there is a large uncertainty of this parameter in the range
$k_{2}f^{2}=2\times 10^{-3} - 10^{-2}$,
as discussed in the next section.}

The pressure gradient force in our isothermal disk model depends solely on the radial density distribution. {In the following} we {take the power-law} disk density distribution {that varies as} $\rho_\mathrm{m} \propto r^{-7/2}$, which is the theoritical density distribution of {steady, isothermal decretion disks (\cite{Por99}; \cite{Oka01}; \cite{Car06})}.

In this section, we {study} the effect of orbital separation {in section \ref{sec:separation}} and 
{that} of binary mass ratio {in section \ref{sec:massratio}}. 
We then briefly discuss the results for 
{B5V stars in section \ref{sec:b5v}} and the effect of {the free} inner boundary condition in section \ref{sec:freebc}.  

\subsection{Effect of binary separation}
\label{sec:separation}

In order to study the effect of binary separation on the oscillation characteristic{s} of the Be disk in {circular binaries}, we vary the value of the binary separation \emph{D} from $D=10R_1$ to $D=50R_1$, with binary mass ratio $q=0.1$.

\begin{figure}
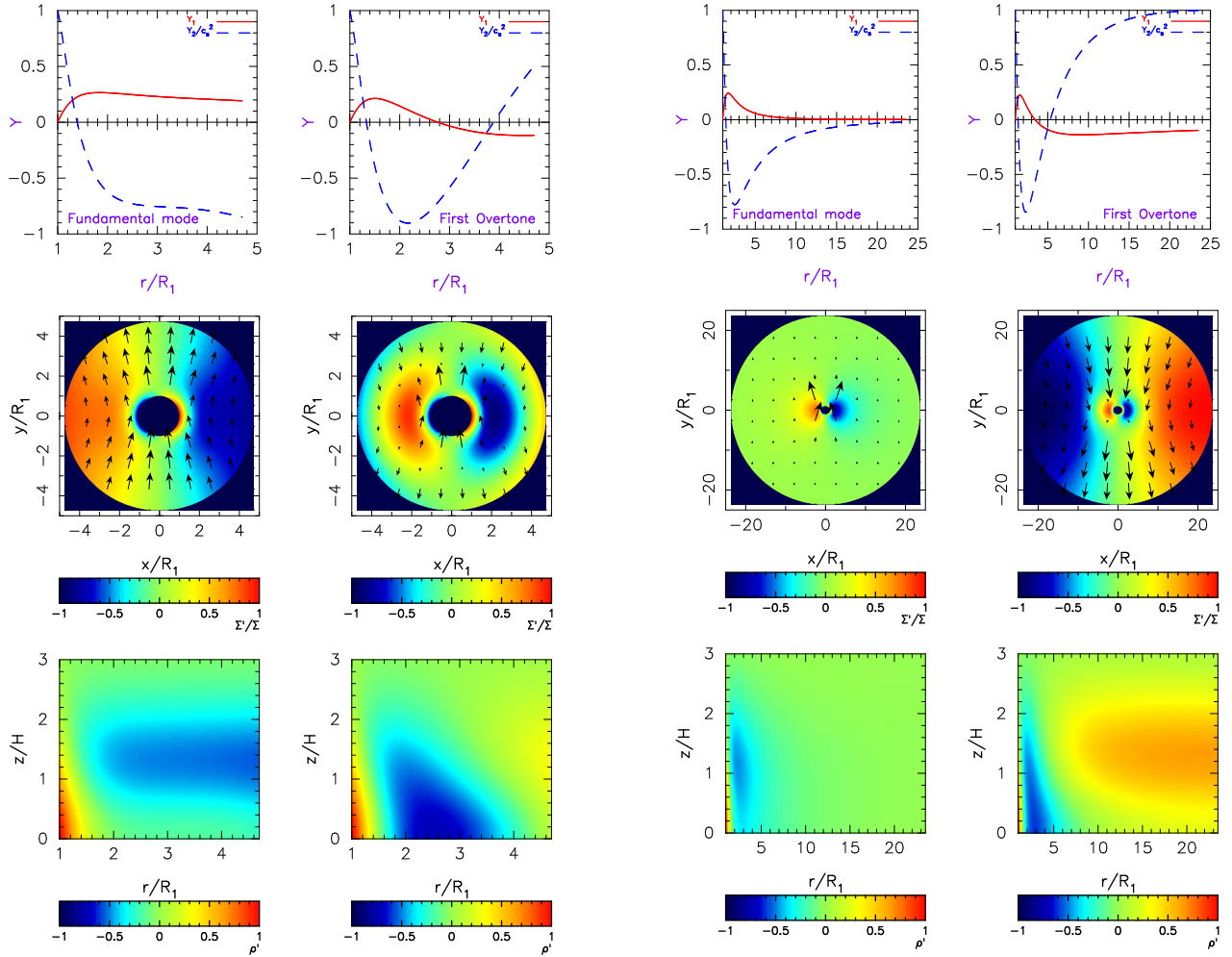

	\centering
         \begin{tabular}{c}
	       \FigureFile(3.5cm,6cm){g1b1k1d1q1ym1.ps}\\
	       \FigureFile(3.5cm,6cm){g1b1k1d1q1dm1.ps}\\
	       \FigureFile(3.5cm,6cm){g1b1k1d1q1tmbd1.ps}
		\end{tabular}
        \begin{tabular}{c}
		  \FigureFile(3.5cm,6cm){g1b1k1d1q1ym2.ps}\\
		  \FigureFile(3.5cm,6cm){g1b1k1d1q1dm2.ps}\\
		  \FigureFile(3.5cm,6cm){g1b1k1d1q1tmbd2.ps}
		   \end{tabular}
		\caption{Eigenfunctions of the $m=1$ fundamental mode (top left) and the first overtone (top right) for $D=10R_1$ and $q=0.1$ with the rigid inner boundary condition. The solid line denotes $Y_{1}$, while the dashed line denotes $Y_{2}/c_{0}^{2}$. In the middle panels, the color-scale plot show the surface density perturbation normalized by the unperturbed  surface density, $\Sigma^{\prime}/\Sigma$, in the $(r,\phi)$-plane, while the arrows denote the perturbed velocity vectors. The bottom panels show the density perturbation, $\rho^{\prime}$, in the $(r,z)$-plane at $\phi = 0$.  }
    \label{fig:dyd10}
\end{figure}

Figures \ref{fig:dyd10} and \ref{fig:dyd50} show the fundamental mode (left) and the first overtone (right) for $D=10R_1$ and $D=50R_1$, respectively. 
The top panels in each figure show the eigenfunction $Y_{1}$ ($=u_{0}$) (solid line) 
and $Y_{2}/c_\mathrm{s}^{2}$ ($=-ih_{0}/c_\mathrm{s}^2=-i\rho_{0}/\rho$) (dashed line) 
in the radial direction. In the middle panels, the color-scale plot shows the surface density perturbation relative to the unperturbed surface density, $\Sigma^{\prime}/\Sigma$, in the $(r,\phi)$-plane: the red (blue) region has a positive (negative) density perturbation. The arrows superposed on the color-scale plot denote the velocity vectors associated with the mode. The disk itself rotates counter-clockwise. The bottom panels shows the density perturbation, $\rho^{\prime}$, in the $(r,z)$-plane of the disk, at $\phi = 0$. Note that the modes are linear, so their normalization is arbitrary.

{All modes in Figs.~\ref{fig:dyd10} and \ref{fig:dyd50}} have positive frequencies, that is, the{y} are prograde modes, which precess in the direction of disk rotation. {Figure~\ref{fig:dyd50} (left) shows that the fundamental mode is well confined in a large disk.} However{, as shown in Fig.~\ref{fig:dyd10} for $D=10R_{1}$,} when the binary separation is small, the confinement of the {fundamental} mode is weak{, so the mode propagates over the whole disk.} As for the first overtones, {the confinement is weak: it is not confined well even for a wide binary with $D=50R_{1}$.}
{Note that these results are consistent with 
\citet{Ogi08} who showed that the 3-D theory allows prograde, fundamental modes
to be confined within several tens of stellar radii in disks around isolated Be stars.}
\begin{figure}
	\centering
        \begin{tabular}{c}
         \FigureFile(3.5cm,6cm){g1b1k1d8q1ym1.ps}  \\
		 \FigureFile(3.5cm,6cm){g1b1k1d8q1dm1.ps} \\
		 \FigureFile(3.5cm,6cm){g1b1k1d8q1tmbd1.ps}
		\end{tabular}
        \begin{tabular}{c}
		  \FigureFile(3.5cm,6cm){g1b1k1d8q1ym2.ps}\\
		  \FigureFile(3.5cm,6cm){g1b1k1d8q1dm2.ps}\\
		  \FigureFile(3.5cm,6cm){g1b1k1d8q1tmbd2.ps}
		\end{tabular}
		\caption{Same as Fig.~\ref{fig:dyd10}, but for $D=50R_1$. }
    \label{fig:dyd50}
\end{figure}

In the middle panels of Figs.~\ref{fig:dyd10} and \ref{fig:dyd50}, we note that perturbation pattern averaged in the vertical direction is similar to that for 2-D disk models around isolated Be stars, which have been extensively studied. We thus expect that the current model can also explain the observed line profile variability. In fact, the modal features suggest that when the violet {(red)} peak is stronger than the red (violet) peak, the line profile as a whole redshifts (blueshifts). This is a typical behavior of the observed line profile variability {(e.g., \cite{Oka91})}.

The bottom panels of Figs.~\ref{fig:dyd10} and \ref{fig:dyd50} show a new feature that has not been suggested by the previous 2-D studies: The density perturbation, $\rho^{\prime} (r,z)$, does not always take the maximum in the disk midplane. The behavior of $\rho^{\prime} (r,z)$ depends on that of the radial component of perturbed velocity, $u_{0}(r)$, as seen from eqauation (\ref{eq:A8}). In regions where $u_{0}$ is negligible, $\rho^{\prime}$ is dominated by the contribution from $h_{0}$, so $\mid \rho^{\prime} \mid $ at a fixed radius has a maximum in the equatorial plane and decreases with vertical coordinate. In contrast, in regions where $u_{0}$ is not negligible, the contribution from $h_{2}$ becomes significant. Then, the density perturbation has a maximum at a vertical coordinate $z \sim (1-2) H$, not in the equatorial plane. In the first overtone, these patterns apear alternately in the radial direction.
 
\begin{figure}
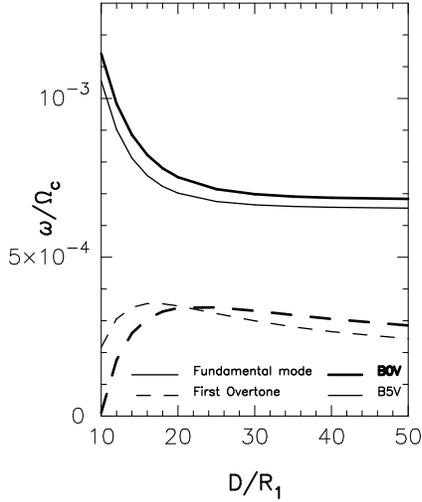

  \begin{center}
    \FigureFile(5.5cm,7cm){g12perbd.ps}
  \end{center}
  \caption{Eigenfrequency as a function of normalized binary separation, $D/R_{1}$, for $q=0.1$. The solid line{s} {denote} the fundamental modes, while the dashed line{s denote} the first overtones. {The thick lines and the thin lines are for a B0V star and a B5V star, respectively. $\omega$ is normalized by the stellar critical rotation frequency $\Omega_\mathrm{c}$.} The rigid boundary condition is applied at the inner disk radius.}
  \label{fig:omega_vs_Dr}
\end{figure}

The distribution of eigenfrequency is shown in Figure~\ref{fig:omega_vs_Dr} as a function of binary separation normalized by the Be star radius, $D/R_{1}$, for $q=0.1$. 
{Here and hereafter, we normalize the eigenfrequency $\omega$ by the stellar critical rotation frequency $\Omega_\mathrm{c}=\sqrt{GM_{1}/R_{1}^{3}}$.}
The solid line{s} and the dashed line{s} denote the fundamental mode{s} and the first overtone{s}, respectively. {The thick lines are for a B0V star
(and the thin lines are for a B5V star discussed in section \ref{sec:b5v})}.
Note that all modes are prograde modes. Note also that the oscillation period of the fundamental mode increases {to an asymptotic value} with increasing binary separation:
{for a B0V star, it} varies from $1.38$\,yr for $D=10R_{1}$ to $2.25$\,yr for $D=50R_{1}$. 
This is because the tidal field is weaker for larger orbital separations, and weaker tidal field decreases the local {apsidal} precession rate $\omega_\mathrm{pr}$
[see eq.~(\ref{eq:omega-pr})]. 
The lower the local precession rate $\omega_\mathrm{pr}$, the lower the eigenfrequency $\omega$. Thus, the eigenfrequency of the {fundamental} mode decreases with increasing orbital separation. However, the effect of the binary separation is only appreciable for small $D/R_{1}$. In systems with large binary separation{, the eigenfrequency} of the fundamental mode changes {little,} because the mode {is} already well confined to the inner part of the disk, so the size of the disk do{es} not matter anymore.  

\subsection{Effect of binary mass ratio}
\label{sec:massratio}

In order to study the effect of binary mass ratio {on disk} oscillation {modes}, we {have} perform{ed} 
{calculations} for $q=0.3${, in addition to the calculations 
for $q=0.1$ shown above}.
We {have} also {carried out calculations for an artificial case
with no tidal effect, except that the disk is truncated at the same radius as for $q=0.1$ 
(hereafter, called the $q=0$ case), 
to compare with the $q=0.1$ result.}

\begin{figure}
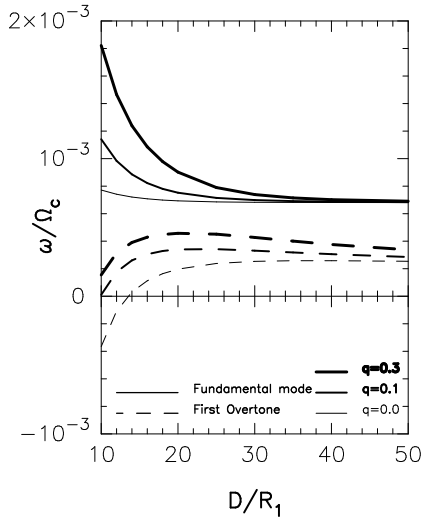

  \begin{center}
    \FigureFile(5.5cm,7cm){g1perbq.ps}
  \end{center}
  \caption{Eigenfrequency as a function of normalized binary separation, $D/R_{1}$, 
  for {$q=0$ (thin lines), 0.1 (lines with intermediate thickness), and 0.3 (thick lines)}. The solid line{s are} for the fundamental mode{s}, while the dashed line{s} for the first overtone{s}. 
  {$\omega$ is normalized by the stellar critical rotation frequency 
  $\Omega_\mathrm{c}$. The rigid inner boundary condition is adopted.}}
  \label{fig:omega_vs_DQr}
\end{figure}

Figure~\ref{fig:omega_vs_DQr} shows the distribution of eigenfrequency as a function of normalized binary separation, $D/R_{1}$, for {$q=0$ (thin lines), 
0.1 (lines with intermediate thickness), and 0.3 (thick lines)}. 
The solid line and the dashed line denote the fundamental mode and the first overtone, respectively. {It is noted that} the eigenfrequency increases with increasing binary mass ratio. This is particularly the case for small $D/R_{1}$. {This is because
t}he local {apsidal} precession rate $\omega_\mathrm{pr}$ increase{s} with increasing mass ratio \emph{q}.
{As mentioned above,} the higher $\omega_\mathrm{pr}${,} the higher {the} eigenfrequency.

Here we compare our result with that of \citet{Ogi08}. \citet{Ogi08} normalized the eigenfrequency $\omega$ and the quadrupole parameter $k_{2}f^{2}$ by $\epsilon^{2}$, where $\epsilon = c_\mathrm{s}(R_{1}/GM_{1})^{1/2}$, which is $\sim 2.3 \times 10^{-2}$ in our case of the B0V star. Then, $k_{2} f^{2} = 5 \times 10^{-3}$ and $\omega/\Omega_\mathrm{c} \sim 6.8 \times 10^{-4}$ for $q=0$ and $D=50R_{1}$ in our case corresponds to $\tilde{Q} \equiv k_{2} f^{2}/\epsilon^{2} \sim 9.1$ and $\tilde{\omega} \equiv \omega/\epsilon^{2} \sim 1.3$, respectively. We note that this is consistent with $\tilde{\omega} \sim 1.2$ for $\tilde{Q} \sim 9$ as seen by eye from Fig.~3 (left) of \citet{Ogi08}).

{Figure~\ref{fig:omega_vs_DQr} suggests that the first overtones are
retrograde modes, which precess in the opposite direction to disk rotation,
for small binary separation ($D \sim 10R_{1}$)  and 
very small mass ratio ($q \ll 0.1$).
Note that the retrograde modes can exist in the Be disk in binaries because of its
truncated outer radius.}

\subsection{Spectral type dependence}
\label{sec:b5v}

As shown by \citet{FH06}, the one-armed oscillation period depend also on the mass and radius of the central star. In this subsection, we discuss the effect of the spectral type. For this purpose, we {calculated the eigenmdoes for} 
a B5V {central} star with $M_{1}=5.9M_{\odot}$, $R_{1}=3.9R_{\odot}$, 
and $T_{\rm eff}=15200 {\rm K}$ \citep{Cox00}.

Figure~\ref{fig:omega_vs_Dr} {compares} the distribution of 
normalized eigenfrequency for 
two different spectral types, B0V (thick lines) and B5V (thin lines). 
{From the figure, we note} that the normalized eigenfrequenc{y} of the fundamental mode for 
{a} B5V star {is slightly} lower than {that} of {a} B0V star.
{This is because both the local apsidal precession rate $\omega_\mathrm{pr}$ and 
the 3-D effect are smaller for a B5V star than for a B0V star: 
A B5V star has a higher critical rotation frequency $\Omega_\mathrm{c}$ and 
a lower sound speed than a B0V star does. This decreases $\omega_\mathrm{pr}/\Omega_\mathrm{c}$.
It also weakens the 3-D effect via a smaller value of $9c_\mathrm{s}^{2}/4r^{2}\Omega$ in eq.~(\ref{eq:basic-2}). The same discussion is applied to the first overtone, except for small binary separation ($D \lesssim 20R_{1}$) where
the tidal effect for a fixed value of $q$ is relatively stronger for a B5V star than for a B0V star.}

{Because of a higher $\Omega_\mathrm{c}$ of a B5V star, 
the dimensional eigenfrequency (or the oscillation period) shows the opposite trend.} Although the normalized eigenfrequency is slightly lower for a B5V star than for a B0V star, the eigenfrequency itself shows the opposite trend. 
Oscillation period  {of} the fundamental mode for a B5V star ranges 
from $0.96$\,yr for $D=10R_{1}$ to $1.5$\,yr for $D=50R_{1}$, 
which is slightly shorter than that {for a} B0V {star} by $\sim 30$\%. 

\subsection{Free inner boundary condition}
\label{sec:freebc}

In long term, some Be stars change {their state} between the Be star phase and the normal B star phase. {There are also other stars that show the loss and reformation of the high-velocity wings of emission lines \citep{Riv01}}. 
Th{ese observational features} indicate that in these stars, the Be disk{, or at least the inner part,} is completely lost and then reformed \citep{Por03}. 
In the formation {stage}, {it is likely that} the Be disk and the central star are {directly} connected. {The rigid inner boundary condition is adequate for such a situation.} {I}n the {dissipation stage, however}, a gap is expected to open between the disk and the central star, {owing to the accretion or ablation of}
the innermost part. 

In order to study {global} oscillation{s} in {dissipating} Be disk{s}, we {have} perform{ed calculations} for the free inner boundary condition, which means that the Lagrangian perturbation of the pressure, $\Delta p$, vanishes at the disk inner radius. For simplicity, we have adopted the same unperturbed disk state as in the rigid boundary case. {Using the same procedure as in section~\ref{sec:bc}, 
the free inner boundary} condition is written as
\begin{equation}
\left(\Omega - \omega\right)Y_{2} -\frac{d\ln\Sigma}{d\ln r} \frac{c_\mathrm{s}^{2}}{r} Y_{1}=0 \mbox{\hspace{1em} at \hspace{1em}} r=R_{1}.
\label{eq:bc-inf}
\end{equation}

\begin{figure}
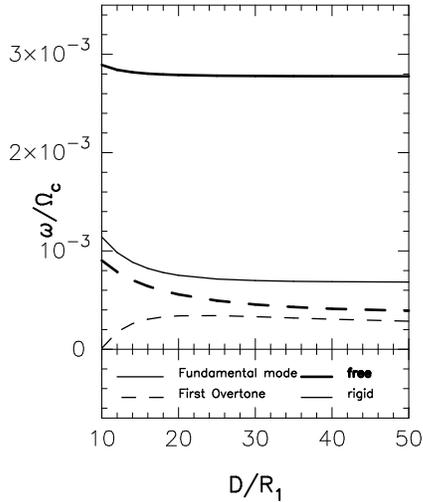

  \begin{center}
    \FigureFile(5.5cm,7cm){ge1perbrf.ps}
  \end{center}
  \caption{Eigenfrequency as a function of normalized binary separation, $D/R_{1}$, for free (thick lines) and rigid (thin lines) inner boundary condition. The solid lines are for the fundamental modes, while the dashed lines for the first overtones.}
  \label{fig:omega_vs_Drf}
\end{figure}

Figure~\ref{fig:omega_vs_Drf} shows the eigenfrequency distribution{s} as a function of normalized binary separation, $D/R_{1}$, for $q=0.1$. {The thick lines are for the free inner boundary condition, 
while the thin lines for the rigid inner boundary condition.} 
The solid lines and the dashed lines denote the fundamental modes and the first overtones, respectively. 
We {note} that {the}
eigenfrequenc{y is} higher {and the mode is more concentrated in the inner part
with the} free {inner} boundary condition 
{than with the rigid one,
as first pointed out by \citet{Pap06} and then confirmed by \citet{Ogi08}. As in the case of rigid inner boundary condition, our result for $q=0.0$ and $D=50R_{1}$ with the free inner boundary condition is consistent with that of \citet{Ogi08}.

\section{Discussion}
\label{sec:discussion}
As mentioned in section~\ref{sec:intro}, binary Be star systems show
two types of long-term V/R variations. The first one is quasi-periodic V/R variations, similar to those observed in isolated Be stars. The second one is periodic V/R variations locked to their orbital period \citep{Ste07}. In our study, we {have} f{ou}nd only eigenmodes compatible with the former type of variations: our eigenmodes depend on the orbital period, but they are not locked to {it}. It is interesting to carry out numerical simulations to study whether tidally forced, nonlinear oscillations can explain the phase locked V/R variations{, but this is beyond the scope of this paper}.

In the previous section, we have shown how global one-armed oscillation modes depend on the binary parameters.
The oscillation period {becomes} longer {in the case of a} wider {binary} separation and/or {a}
lower binary mass ratio.
Such dependence, however, will be observed only for short period binaries.
In wide binaries where the disk size is large enough to confine the one-armed modes
in the inner part of the disk,
the mode characteristics depend little on the binary parameters.
The critical orbital separation{, where the size of the Be disk is just large enough to confine
the fundamental mode,} depends on the details of the model{.}
{The difference between the model oscillation periods (a few yr) and the observed ones
(5-10 yr) implies that there are still missing mechanisms that affect the global disk oscillations.
One of them could be the optically thick line forces examined by \citet{Gay01}, who
found that these forces have a net effect that likely lowers the eigenfrequency of the one-armed mode.
The more realistic density and temperature distributions could also contribute to lower the eigenfrequency.}
{However, despite these mechanisms not taken into account,} 
we believe that the current results are qualitatively robust.

{Observationally, no clear dependence of the V/R period on the binary parameters
has been found. \citet{Rei05} compared the time-scales of disk formation/dissipation
cycle and V/R variability with the orbital period for eight Be/X-ray binaries (see their Table~3). 
They found a good correlation
between the time-scale of disk formation/dissipation cycle and the orbital period,
which agrees with the scenario that the Be disk is truncated by the tidal torques from the neutron star \citep{One01}. 
On the other hand, the V/R variability time-scale has no clear correlation
with the orbital period, although short period systems have, in general, 
short V/R time-scale, with which our}
results qualitatively agree. 

In {this paper}, we {have} assume{d} an inviscid disk for mathematical simplicity. However, viscosity in the Be disk is considered to play an important role in the formation of a nearly Keplerian disk by causing the outward
drift of matter ejected from {the} equatorial surface of {the} Be star, as proposed in the viscous decretion disk model by \citet{Lee91}. 
\citet{Neg01} show{ed, after  \citet{Kat78},} that global $m=1$ modes, which are neutral in inviscid disks, become overstable when the visco{us} effect is {taken into account as a perturbation}. 
The growth rate of the mode, i.e., the imaginary part of the eigenfrequency, {of the mode} is proportional to {the Shakura-Sunyaev viscosity parameter $\alpha$, while the real part is independent of $\alpha$ 
as long as $\alpha \ll 1$. \citet{Ogi08} also obtained the result that the real part of an eigenfrequency in a viscous disk with $\alpha=0.1$ is only slightly different from that in an inviscid disk.}
Thus, {we expect that as far as the oscillation period is concerned,} our result {is valid}
for viscous disks {with small $\alpha$} as well.

{Finally, we briefly comment on the effect of stellar rotation 
via the quadrupole parameter $k_{2}f^{2}$.
For B-type main sequence stars,
the stellar evolution calculations result in the apsidal motion constant $k_2$ in the range
$2.5 \times 10^{-3} \le k_2 \le 10^{-2}$, depending on the evolutionary stage and
the internal angular momentum distribution (\cite{CG91}; \cite{Cla95}; \cite{Cla99}).
While it remains controversial how close the rotation of Be stars {is} to 
the critical rotation, it is likely $f=\Omega_{1}/\Omega_\mathrm{c} \gtrsim 0.9$ 
(e.g., \cite{Fre05}), {which} yield{s} for the quadrupole factor in the range
$2 \times 10^{-3} \lesssim k_{2}f^{2} \lesssim 10^{-2}$.}

{In order to study the effect, we have calculated the eigenmodes for 
$k_{2}f^{2} = 10^{-2}$, implicitly assuming an extremely high 
value of $k_{2} \sim 10^{-2}$ for a critically rotating star ($f \sim 1$),
for a B0V star in a binary with $q=0.1$.}
{As expected, the resulting eigenfrequency was higher than that for $k_{2}f^{2} = 5 \times 10^{-3}$: It increased from $\omega \sim 1.1 \times 10^{-3}$ to $\omega \sim 1.6 \times 10^{-3}$ for $D=10R_{1}$
and from $\omega \sim 6.8 \times 10^{-4}$ to $\omega \sim 1.3 \times 10^{-3}$ for $D=50R_{1}$.}
This is because the local {apsidal} precession rate $\omega_\mathrm{pr}$ 
{appreciablly increases} in the inner part {with an increase in $k_{2}f^{2}$}.
For the first overtones, we also found higher eigenfrequenc{ies} for {the larger quadrupole parameter.}  However, the {difference is much smaller than in the case of
fundamental modes, in particular} for systems with large orbital separations.

\section{Conclusions}
We have studied the tidal effect of the companion on the global oscillation modes in equatorial disks around binary Be stars. For simplicity, we assumed {the binary orbit to be circular and} the {Be} disk to be inviscid, isothermal, and
truncated at the tidal radius. 
We {solved} linearized {equations for global} $m=1$ 
{perturbations in a}
three-dimensional {Be} disk {with a power-law density distribution,} 
with the rigid wall inner boundary condition, which is applicable to systems where material is ejected continuously, so there is no gap between the star and the disk.

We {have} obtained
prograde {fundamental} modes {even when the quadrapole contribution to the potential is negligible.} This confirms the results of \citet{Ogi08}}.
In our study of binary Be star systems, the modes are well confined when the disk is larger than a few tens of stellar radii. 

We have found that the oscillation period increases
with increasing binary separation {and/or {decreasing} binary mass ratio.
The effect is, however, only appreciable for small binary separation. 
In systems with large binary separations, 
the fundamental mode is well confined to the inner part of the disk, 
so the eigenfrequency no longer depends on the binary parameters.

Be stars sometime{s} show {a} cavity {between the star and} 
the disk when the disk is being dissipated.
{In order to study the global oscillation modes in such disks,
we have solved the perturbation equations with a free inner boundary condition,
without changing the density distribution.
We have obtained much higher eigenfrequencies with the free inner bounday condition
than with the rigid one,
as in \citet{Pap06} and \citet{Ogi08}.
The result implies an interesting possibility that 
the V/R variability time-scale is shorter when the disk is dissipating
than when it is forming or persistent.}

{In this paper we have assumed a power-law density distribution.
In the disk dissipation stage, however, it is likely that} the density distribution is 
far from {a} power-law {form} and {is} highly peaked 
near the inner {radius}. 
In a subsequent paper, we will study {the} effect {of density distribution on
the global $m=1$ oscillations}.

\bigskip

FO thank{s} Masayuki Fujimoto for helpful discussions. She also acknowledges the scholarship from Ministry of Education, Culture, Sports, Science and Technology.
{We thank Gordon Ogilvie for kindly showing us his latest result
on the three-dimensional effect on the $m=1$ mode confinement.}


\begin{thebibliography}{}

\bibitem[Carciofi \& Bjorkman(2006)]{Car06}
{Carciofi, A.C., \& Bjorkman, J.E., 2006, ApJ, 639, 1081}

\bibitem[Castor et al.(1975)]{CAK75}
{Castor, J. I., Abbott, D. C., \& Klein, R.I. 1975, \apj, 195,
157}

\bibitem[Chen \& Marlborough(1994)]{CM94}
{Chen, H. \&, Marlborough, J.M. 1994, \apj, 427, 1005}

\bibitem[Claret(1995)]{Cla95}
{Claret, A. 1995, \aaps, 109, 441}

\bibitem[Claret(1999)]{Cla99}
{Claret, A. 1999, \aap, 350, 56}

\bibitem[Claret \& Gimnez(1991)]{CG91}
{Claret, A. \&, Gimnez, A. 1991, \aaps, 87, 507}

\bibitem[Collins(1987)]{Col87}
Collins II, G.W.\ 1987, in IAU Colloq. 92, Physics of Be Stars,
   ed. A.Slettebak \& T.P. Snow (Cambridge Univ. Press,
   Cambridge) p.3
   
\bibitem[Cox(2000)]{Cox00}
Cox, A.N.\ 2000, in Allen`s Astrophysical Quantities Fourth Edition (The Athlone Press, London), p.389

\bibitem[Eggleton(1983)]{Egg83}
{Eggleton, P. P. 1983, \apj, 268, 368}

\bibitem[Fr\'{e}mat et al.(2005)]{Fre05}
{Fr\'{e}mat, Y., Zorec, J., Hubert, A.-M., \& Floquet, M. 2005, \aap, 440, 305}

\bibitem[Fi\v{r}t \& Harmanec(2006)]{FH06}
{Fi\v{r}t, R. \& Harmanec, P. 2006, \aap, 447, 277}

\bibitem[Friend \& Abbott(1986)]{FO86}
{Friend, D. B. \& Abbott, D. C. 1986, \apj, 311, 701}

\bibitem[Gayley et al.(2001)]{Gay01}
{Gayley, K. G., Ignace, R.., Owocki, S. P. 2001, \apj, 558, 802}

\bibitem[Hirose \& Osaki(1993)]{Hir93}
Hirose, M., Osaki, Y.\ 1993, \pasj, 45, 595

\bibitem[Kato (1983)]{Kat83}
Kato, S. \ 1983, \pasj, 35, 249

\bibitem[Kato et al.(1978)]{Kat78}
Kato, S.\ 1978, \mnras, 185, 629

\bibitem[Lee et al.(1991)]{Lee91}
Lee, U., {Saio, H., Osaki, Y.}\ 1991, \mnras, 250, 432

\bibitem[Negueruela et al.(2001)]{Neg01}
{Negueruela I., Okazaki A.T., Fabregat J., Coe M.J., Munari U., Tomov
  T., 2001, \aap, 369, 117}

\bibitem[Ogilvie(2008)]{Ogi08}
Ogilvie, G.I. \ 2008, \mnras, {in press}

\bibitem[Okazaki(1991)]{Oka91}
Okazaki, A.T.\ 1991, \pasj, 43, 75

\bibitem[Okazaki(1997)]{Oka97}
Okazaki, A.T.\ 1997, \aap, 318, 548

\bibitem[Okazaki(2001)]{Oka01}
{Okazaki, A.T.\ 2001, \pasj, 53, 119}

\bibitem[Okazaki et al.(1987)]{Oka87}
{Okazaki, A. T., Kato, S, Fukue, J. 1987, \pasj, 39, 457}

\bibitem[Okazaki \& Negueruela(2001)]{One01}
Okazaki, A.T., Negueruela, I. \ 2001, \aap, 369, 108

\bibitem[Papaloizou et al.(1992)]{Pap92}
Papaloizou, J.C., Savonije, G.J., Henrichs, H.F.\ 1992, \aap, 265, L45

\bibitem[Papaloizou \& Savonije(2006)]{Pap06}
Papaloizou, J.C.B, Savonije, G.J.\ 2006, \aap, 456, 1097

\bibitem[Porter(1999)]{Por99}
{Porter, J.M. 199, \aap, 348, 512}

\bibitem[Porter \& Rivinius(2003)]{Por03}
Porter, J.M., Rivinius, T. \ 2003, \pasp, 115, 1153

\bibitem[Reig et al.(2005)]{Rei05}
Reig, P., Negueruela, I., Fabregat, J., Chato, R., Coe, M.J.\ 2005,
   \aap, 440, 1079

\bibitem[Rivinius et al.(2001)]{Riv01}
{Rivinius, Th., Baade, D., \v{S}tefl, S., Maintz, M. 2001, \aap,
  379, 257}

\bibitem[Savonije \& Heemskerk(1993)]{Sav93}
Savonije, G.J., Heemskerk, M.H.M.\ 1993, \aap, 276, 409

\bibitem[\v{S}tefl et al.(2007)]{Ste07}
\v{S}tefl, S., Okazaki, A.T., Rivinius, T., Baade, D. \ 2007, in ASP
   Conf. Series, Active OB Stars: Laboratories for Stellar \&
   Circumstellar Physics (Astronomical Society of the Pacific, San Francisco), p.274

\bibitem[Telting et al.(1994)]{Tel94}
Telting, J.H., Heemskerk, M.H.M, Henrichs, H.F., Savonije, G.J. 
   1994, \aap, 288, 558

\bibitem[Vakili et al.(1998)]{Vak98}
{Vakili, F., Mourard, D., Stee, PH., Bonneau, D., Berio, P.,
   Chesneau, O., Thureau, N., Morand, F., Labeyrie, A., \&
   Tallon-Bosc, I. 1998, \aap, 335, 261}

\bibitem[Whitehurst \& King(1991)]{WK91}
{Whitefurst, R., King, A. 1991, \mnras, 249, 25}

\end{thebibliography}
\end{document}